# Band gap engineering by functionalization of BN sheet

Authors: A. Bhattacharya[1], S. Bhattacharya[1], G. P. Das[1*]

Department of Materials science, Indian Association for the Cultivation of Science, Jadavpur, Kolkata 700032, INDIA.

[*]Corresponding Author's email: msgpd@iacs.res.in

Authors' email address: msab2@iacs.res.in, mssb3@iacs.res.in

**Abstract:**

From first principles calculations, we investigate the stability and physical properties of single layer *h*-BN sheet chemically functionalized by various groups viz. H, F, OH, $CH_3$, CHO, CN, $NH_2$ etc. We find that full functionalization of h-BN sheet with these groups lead to decrease in its electronic band gap, albeit to different magnitudes varying from 0.3 eV to 3.1 eV, depending upon the dopant group. Functionalization by CHO group, in particular, leads to a sharp decrease in the electronic band gap of the pristine BN sheet to ~ 0.3 eV, which is congenial for its usage in transistor based devices. The phonon calculations on these sheets show that frequencies corresponding to all their vibrational modes are real (positive), thereby suggesting their inherent stability. The chemisorption energies of these groups to the B and N atoms of the sheet are found to lie in the range of 1.5 -6 eV.

**Manuscript Text:**

Graphene[1], the two dimensional $sp^2$-bonded single layer of graphite, is a hot pursuit today because of its unique combination of electrical[2] and mechanical[3] properties. The conductivity of graphene being very high (its electrons acting as massless fermions), its usage in electronic devices can be made possible only if one can somehow introduce a band gap (close to those of conventional semiconductors such as Si, GaAs etc) in it. Chemical modification of graphene by functionalizing its surface with various dopants, provide one such avenue for opening of a band gap and has been found to be useful in band gap engineering of graphene[4]. In fact, the fully hydrogenated derivative of graphene, known as graphane[5-7] has been experimentally synthesized in laboratory and is found to have an insulating band gap of ~3.5 eV. Another analogous two dimensional nanostructure viz. hexagonal Boron Nitride sheet (*h*-BN sheeet)[8] has emerged as an strong candidate in this field because of its modified electronic properties but similar hexagonal planar geometry close to graphene. *h*-BN sheet has been experimentally synthesized in single and multiple layers.[9,10] While graphene is a semi-metal with unique 2D conducting properties, *h*-BN is a typical insulator. Hydrogenation of *h*-BN sheet leads to reduction in the band gap of the sheet, opposite to the trend observed in case of graphene.[6,11] Thus, chemical modification seems to be a natural tool for tuning the gap in both graphene and BN sheet. For both graphene and BN sheet, H passivation takes place in such a way that one H atom gets bonded to each of the C or B and N atoms of the sheet in specific periodic fashion, giving rise to various possible structural conformers of the sheet viz. chair, boat and stirrup.[6,11,12] In the chair and boat conformers, H atom alternates singly and in pair on both sides of the sheet,[5,6,11] while in the stirrup conformer, three consecutive H atoms alternate on either side of the sheet.[6,11] In graphane, the chair conformer has been found to have the highest stability.[6] However, in case of hydrogenated *h*-BN sheet, the stirrup conformer is found to have the highest stability (binding energy of ~ 4.84 eV/atom),[11] followed by boat and chair conformers. Though chemical modification of graphene has been studied and experimented by several groups, the same in *h* BN-sheet has not been explored exhaustively till date.

In this present work, we estimate from our first principles based calculations, the stability and ground state properties of *h*-BN sheet chemically functionalized by various groups viz. H, F, OH, $CH_3$, CHO, CN, $NH_2$ etc. Using these dopants, the band gap of these chemically modified sheets can be tuned from 3.2 eV to 0.3 eV. The phonon calculations can play an important in order to determine the inherent stability of these sheets. Therefore, we have also plotted the phonon dispersion and density of states plots to show that frequencies corresponding to all the vibrational modes in these sheets are real (positive) thereby suggesting their inherent stability.

Our calculations have been carried out using first-principles density functional theory (DFT)[13,14] based on total energy calculations within the local density approximation (LDA). LDA rather than GGA is found to yield better band gap and related results for two-dimensional systems with week interlayer binding[15]. We have used VASP[16] code with projected augmented wave (PAW) potential[17] for all elemental constituents of these functionalized sheets. The calculations have been performed using the CA exchange correlation functional[18]. An energy cut off of 600eV has been used. The k-mesh was generated by Monkhorst–Pack[19] method and the results were tested for convergence with respect to mesh size. In all our calculations, self-consistency has been achieved with a 0.0001 eV convergence in total energy. For optimizing the ground state geometry[20, 21], atomic forces were converged to less than 0.001eV/Å via conjugate gradient minimization. The Mulliken population analysis[22] and phonon calculations have been carried out by using CASTEP code[23] where the LDA calculations are performed by CA-PZ exchange-correlation. A plane wave basis set with norm conserving pseudo potential has been used for the phonon calculations.

In *h*-BN sheet, each B (N) atom is bonded to three N (B) atoms giving rise to repetitive hexagonal planar layered structure where each B-N bond has a bond length of 1.45 Å. Our calculations on single layer *h*-BN sheet, show that it has a band gap of ~ 4.5 eV [Fig-1(a)] and high stability with an estimated binding energy (BE) of ~7 eV/atom. The phonon density of states and dispersion of *h*-BN sheet is given in Fig-2(a). The phonon spectra of *h*-BN sheet cover the frequency range of 0 to ~1500 cm$^{-1}$. The B-N bond shows highest phonon density of states at a frequency of 1360 cm$^{-1}$ which is both IR and Raman active. This is in conformity with the experimental results of Gorbachev *et al* on monolayer *h*-BN sheet[25]. We have studied the effect of chemical modification on the electronic structure of single layer *h*-BN sheet by functionalizing it with various atoms / groups of atoms on both sides of the sheet. The mono-atomic groups, considered in our study, are hydrogen and fluorine, while the other groups can be categorized according to their atomic composition as (a) Oxygen containing groups (viz. OH, CHO, COOH, $H_2O$), (b) Nitrogen containing groups (viz. CN, $NH_2$) and (c) hydrocarbon group (viz. $CH_3$). Out of these groups, *h*-BN sheet does *not* bind $H_2O$ and COOH (carboxyl group) molecules, while it chemisorbs all the other groups mentioned above with binding energy ranging from ~1.5 to 6 eV, as enlisted in Table-1. Full functionalization of the sheet takes place in a way that one group gets bonded to each of the B and N atoms, giving rise to various structural conformations viz. chair, boat and stirrup as explained in the introduction. A hexagonal unit of the *h*-BN sheet is shown in the inset of Fig-1(a). The B and N atoms in the hexagon are marked from 1 to 6. In chair conformer, the groups chemisorbed at 1, 3, 5 sites point up while at 2, 4, 6 sites point down the BN plane. In boat and stirrup conformers, the groups pointing up are chemisorbed at 1, 2, 4, 5 and 1, 2, 3 sites of the hexagon respectively while all the rest sites have groups chemisorbed below the BN plane. We find that other than hydrogenation of BN sheet (where the stirrup conformation is found to be energetically most stable conformer by ~0.2 eV/atom), in all other cases the chair conformation shows highest stability. This is due to the increase in size of the groups (as one goes from H to $CH_3$) which gives rise to high repulsive interaction between the groups chemisorbed on consecutive sites in boat and stirrup conformers. In chair conformer, the groups bonded to two consecutive sites, face opposite sides of the BN plane (1-up, 1-down geometry) and thus, the distance between two functionalizing groups in any plane is always higher than that in boat and stirrup conformers. Therefore, the repulsive interaction between the chemisorbed groups is lowest in chair conformers (excepting hydrogenation of BN sheet). So, we restrict our calculations to the chair conformation of these chemically modified *h*-BN sheets and discuss on their stability and physical properties.

The electronic properties and application of **hydrogenated** *h*-BN sheet has been studied by various groups recently.[11,26,27] A fully hydrogenated *h*-BN sheet (BHNH sheet) is found to have BE of ~ 4.6 eV/atom. Fig-1(b) shows the structure (of single unit cell), electronic density of states (DOS) and band structure plots of the chair BHNH sheet. From the site projected DOS plot, hybridization between the B, N and H atoms of the sheet can be seen. The sheet shows a direct band gap of ~ 3.0 eV at the Γ-point. The phonon spectrum of the hydrogenated sheet shows vibrational frequency modes distributed from 0 to ~ 3250 $cm^{-1}$ which indicates the inherent stability of the system [Fig-2(b)]. The acoustics bands (ranging from 0-600 $cm^{-1}$) are comprised of the vibrational modes corresponding to the B-N bonds of the sheet and are distinctly separated from the opical bands (800-3250 $cm^{-1}$). The phonon dispersion shows two sharp optical bands at 2600 $cm^{-1}$ and 3245 $cm^{-1}$ corresponding to vibrational modes of B-H and N-H bonds respectively. The highest vibrational frequency modes corresponding to the B-H, B-N and N-H bonds in the sheet are enlisted in Table-1. Though the B atoms are lighter than the N atoms, the highest vibrational frequency mode corresponding to B-H bond is found to be lower than that of the N-H bond. This is because of the higher chemisorption energy of H (and hence greater bonding) to the B atoms (4.3eV/H) as compared to the N atoms (3.9 eV/H) of the sheet.

A fully **fluorinated BN sheet** (BFNF sheet) is found to have direct energy band gap of 3.1 eV at the Γ-point [Fig-1(c)]. It has stability (BE) of ~ 4.85 eV/atom which is higher than that of its hydrogenated counter-part. The F atoms are chemisorbed with binding energy of 5.8 eV/F and 3.5 eV/F to each of the B and N atoms of the sheet respectively. In order to establish the stability of the BFNF sheet, we have carried out phonon calculations and the corresponding phonon dispersion and density of states are shown in Fig-2(c). All the vibrational frequency modes in the BFNF sheet are found to be real positive, ranging from 0-1500 $cm^{-1}$. Thus, vibrational modes in BFNF sheet are much lower in frequency as compared to those in BHNH sheet which is due to higher mass of the F atoms in the fluorinated sheet. The acoustic and optical bands can not be clearly separated. However, the last band of optical spectrum is separated from the lower bands and corresponds to vibrational modes corresponding to the B-F bond (1300 cm$^{-1}$). The highest vibrational modes corresponding to the B-F bond (1300 cm$^{-1}$) is higher than that of the N-F bond (1050 cm$^{-1}$) which is due to the lighter mass of B atoms.

Functionalization of BN sheet with **alcohol (OH) group** also leads to decrease in the band gap of the sheet. The structure, electronic DOS and band structure plots of BOHNOH sheet are shown in Fig-1 (d). It has stability (in terms of BE) of 4.6 eV/atom and also shows a direct energy band gap of 2.3 eV at the Γ-point. The O-H bond subtends oblique angles to the B (110$^o$) and N (105$^o$), as shown in the inset of Fig-1(d). The chemisorption energy of OH group to the N atoms (2.47 eV/group) is estimated to be higher than that of the same with B atoms (1.57 eV/group) and therefore, the highest vibrational frequency corresponding to B-O$_B$ bond (1240 cm$^{-1}$) is found to be higher than that of N-O$_N$ bond (1100 cm$^{-1}$) as shown in Table-1 (O$_B$ and O$_N$ are the symbols used for oxygen bonded to the B and N atoms of the sheet respectively). The phonon dispersion of the BOHNOH sheet [Fig-2(d)] shows that the vibrational frequency modes are distributed from 0 to 3400 cm$^{-1}$. The optical spectra (ranging from 600 to 3400 cm$^{-1}$) show a number of separated groups of phonons, out of which the highest frequency bands corresponds to the vibration of O$_B$-H (at 3430 cm$^{-1}$) and O$_N$-H bonds (at 3350 cm$^{-1}$).

Chemical modification of *h*-BN sheet with **aldehyde (CHO) group** is very crucial in this study as it leads to huge lowering in the electronic band gap of the sheet (to ~ 0.3 eV). The structural unit, electronic band dispersion and DOS plots of BCHONCHO sheet are given in Fig-1(e). It has stability (BE) of 4.9 eV/atom. The band structure is mainly modified due to the lowering in unoccupied bands corresponding to C and H atoms near the M-point which gives rise to an indirect band gap of the sheet. The highest vibrational frequency corresponding to B-C$_B$ (1000 cm$^{-1}$) and N-C$_N$ (1000 cm$^{-1}$) bonds are close in the BCHONCHO sheet (Table-1). This is because though B is lighter but the chemisorption energy of the CHO group to B atoms (2.3 eV/group) is higher as compared to that to the N atoms (2.1 eV/group) of the sheet (Table-1). The corresponding phonon dispersion and DOS plots are shown in Fig-2(e). The vibrational modes are ranging from 0 to 3500 cm$^{-1}$. The optical spectrum (600- 3500 cm$^{-1}$) is comprised of many separated single bands of phonon. The highest optical frequency bands at 3350 cm$^{-1}$ and 2750 cm$^{-1}$ corresponds to the vibration of the $C_N$-H and $C_B$-H bonds. However, the highest vibrational modes of the $C_B$-O and $C_N$-O bonds corresponds to a frequency of 1800 cm$^{-1}$ which gives rise to two overlapping optical bands at this frequency.

Chemical modification of the BN sheet by **methyl (CH$_3$) group** alters the electronic band gap of the sheet to an intermediate value as reached by functionalization of the sheet with H, F, OH and CHO groups. The BCH$_3$NCH$_3$ sheet has a BE of ~ 4.5 eV/atom and is found to be a direct band gap semiconductor with an energy gap of 1.1 eV at the Γ-point. The corresponding band structure and DOS plots are given in Fig-1(f). In the occupied region of the DOS, hybridization between the H, B, C and N atoms of the sheet can be seen. The phonon dispersion and phonon DOS of the BCH$_3$NCH$_3$ sheet is given in Fig-2(f). The acoustic and optical regions merge with each other in the phonon spectra. In the higher part of optical spectrum, six sharp optical bands can be seen from 3000 cm$^{-1}$ to 3600 cm$^{-1}$, which corresponds to the highest vibrational modes of the $C_B$-H and $C_N$-H bonds. The chemisorption energy (Table-1) of CH$_3$ group to the B atom (2.32 eV/CH$_3$) is higher than that of the same to the N atom (1.81 eV/CH$_3$) which also reflects in the highest vibrational frequency corresponding to the B-$C_B$ (1270 cm$^{-1}$) and N-$C_N$ (1467 cm$^{-1}$) bonds.

Chemical functionalization of the sheet with groups containing N viz. **CN** and **NH$_2$**, decrease the electronic band gap of the corresponding sheets to 2.2 eV (indirect) and 2.5 eV (direct-Γ) respectively. The band structure and DOS plots of BCNNCN and BNH$_2$NNH$_2$ sheets are shown in Fig-1(g) and Fig-1(h) respectively. While the stability of the BCNNCN sheet (BE of 5.74 eV/atom) is found to be the highest, the stability of the BNH$_2$NNH$_2$ sheet (BE of 4.03 eV/atom) is found to be the lowest amongst all other chemically modified sheet. The phonon spectrum of BCNNCN sheet is shown in Fig-2(g). All the vibrational frequency modes in the sheet are found to be positive. The phonon spectrum also show separated groups of phonons. Two sharp optical bands can be seen at frequency of 2750 cm$^{-1}$, which corresponds to the highest vibrational modes of both $C_B$-N and $C_N$-N bonds. The highest vibrational frequency modes corresponding to the B-$C_B$ (977 cm$^{-1}$) and N-$C_N$ (1000 cm$^{-1}$) bonds are close since the chemisorption energy of the CN group to B (4.5 eV/CN) is estimated to be higher than that to N (3.2 eV/CN) as given in Table-1. The acoustic and optical bands merge in the vibrational spectrum of BNH$_2$NNH$_2$ sheet as shown in Fig-2(h). There are four sharp optical bands at frequency range of 3000 cm$^{-1}$ to 3600 cm$^{-1}$, corresponding to the vibrational modes of the N$_B$-H and N$_N$-H bonds. The chemisorption energy of NH$_2$ group to B atoms (2.56 eV/NH$_2$) is found to be higher than that to N atoms (2.19 eV/NH$_2$) atoms of the sheet, which also reflects, in the vibrational frequency of the B-N$_B$ and N-N$_N$ bonds (Table-1).

From the Mulliken population analysis of these chemically functionalized sheets and native *h*-BN sheet (see Table-2), the charge state of B and N atoms before and after functionalization can be compared. Native *h*-BN sheet shows a Mulliken charge transfer of +0.42 electrons from B to N atoms of the sheet. However, in most of the functionalized sheets, N gains electron and acquires a more negative charge state while B loses electron to acquire a more positive charge state as compared to their charge state in native *h*-BN sheet. This is due to the electronegativity difference of the atoms of the sheet and dopant groups. The electronegativity of atoms present in the dopant groups (viz. O in OH, C in CHO, CN, CH$_3$ and N in NH$_2$, via which the bonding takes place) is lower than their native electronegaitivity, due to transfer of electrons. Therefore, in most of the cases, the electronegativity of the dopant atom via which the bonding with the sheet takes place, are found to be higher than B but lower than N atoms of the sheet. Thus, the dopant groups being more electronegative than B gains electron from it while lose electrons to the N atoms. However, the fluorinated BN sheet is an exception to this generalization. In BFNF sheet, both B and N lose electron. The reason can be explained from the Pauling electronegativity difference of native B (2.04), N (3.04) and F (3.98) atoms. The electronegativity of F is higher than both B and N atoms and therefore in BFNF sheet, both B and N lose electron to F atoms. Thus, in all these functionalized sheets, the dopant atoms undergo charge transfer with the atoms of the sheet which in turn alter their electronic structure.

In summary, we have performed first principles calculations to estimate the stability and ground state properties of *h*-BN sheet chemically functionalized by various groups viz. H, F, OH, CH$_3$ CHO, CN, NH$_2$ etc. Using these dopants, the band gap of these chemically modified sheets can be tuned from 3.2 eV to 0.3 eV. Most of these functional groups, excepting CHO and CN, results in direct band gap semiconductors. Functionalization by CHO group, in particular, leads to a sharp decrease in the electronic band gap of the pristine BN sheet to ~0.3 eV, which is congenial for its usage in transistor based devices. We have also performed phonon calculations on these functionalized sheets to show that the frequencies corresponding to all their vibrational modes are real (positive) suggesting their inherent stability. The chemisorption energy of these groups to the B and N atoms of the sheet are found to lie in the range of 1.5 - 6 eV.

Table-1: Structural and ground state properties of BN sheet functionalized by various groups (The subscripts in second column denote the host element to which the dopant atom is bonded)

| Systems | Bond | Length (Å) | Highest Vib. Freq. mode (cm$^{-1}$) | LDA Band gap (eV) | BE of the system (eV/atom) | Chemisorption energy of group bonded to (eV/group) | |
|---|---|---|---|---|---|---|---|
| | | | | | | B | N |
| BN | B-N | 1.45 | 1360 | 4.4 | 7.043 | - | - |
| BHNH | B-H | 1.20 | 2600 | 3.0 (direct-Γ) | 4.574 | 4.30 | 3.90 |
| | B-N | 1.58 | 830 | | | | |
| | N-H | 1.03 | 3245 | | | | |
| BFNF | B-F | 1.35 | 1300 | 3.1 (direct-Γ) | 4.850 | 5.84 | 3.52 |
| | B-N | 1.62 | 820 | | | | |
| | N-F | 1.44 | 1050 | | | | |
| BOHNOH | O$_B$-H | 0.98 | 3430 | 2.3 (direct-Γ) | 4.663 | 1.57 | 2.47 |
| | B-O$_B$ | 1.40 | 1240 | | | | |
| | B-N | 1.63 | 910 | | | | |
| | N-O$_N$ | 1.48 | 1100 | | | | |
| | O$_N$-H | 0.99 | 3350 | | | | |
| BCNNCN | C$_B$-N | 1.16 | 2270 | 2.2 | 5.746 | 4.52 | 3.12 |
| | B-C$_B$ | 1.52 | 977 | | | | |
| | B-N | 1.73 | 790 | | | | |
| | N-C$_N$ | 1.40 | 1000 | | | | |
| | C$_N$-N | 1.16 | 2270 | | | | |
| BCH$_3$NCH$_3$ | C$_B$-H | 1.06 | 3390 | 1.1 (direct-Γ) | 4.541 | 2.32 | 1.82 |
| | B-C$_B$ | 1.59 | 1270 | | | | |
| | B-N | 1.67 | 990 | | | | |
| | N-C$_N$ | 1.49 | 1467 | | | | |
| | C$_N$-H | 1.05 | 3520 | | | | |
| BNH$_2$NNH$_2$ | N$_B$-H | 1.01 | 3544 | 2.5 (direct-Γ) | 4.028 | 2.58 | 2.19 |
| | N$_B$-H | 1.01 | 3510 | | | | |
| | B-N$_B$ | 1.48 | 1240 | | | | |
| | B-N | 1.74 | 800 | | | | |
| | N-N$_N$ | 1.49 | 1300 | | | | |
| | N$_N$-H | 1.01 | 3386 | | | | |
| | N$_N$-H | 1.02 | 3315 | | | | |
| BCHONCHO | C$_B$-H | 1.11 | 2760 | 0.3 | 4.9016 | 2.3 | 2.1 |
| | C$_B$-O | 1.20 | 1800 | | | | |
| | B-C$_B$ | 1.61 | 1000 | | | | |
| | B-N | 1.75 | 790 | | | | |
| | N-C$_N$ | 1.50 | 1000 | | | | |
| | C$_N$-O | 1.19 | 1800 | | | | |
| | C$_N$-H | 1.06 | 3360 | | | | |

Table-2: Mulliken population analysis of BN sheet and it functionalized derivatives.

| System | Atom | Mulliken charge | System | Atom | Mulliken charge |
|---|---|---|---|---|---|
| **BN** | B | 0.420 | **B(CH$_3$)N(CH$_3$)** | H$_{CB}$ | 0.150 |
| | N | -0.420 | | C$_B$ | -0.593 |
| **BHNH** | H$_B$ | -0.033 | | B | 0.709 |
| | B | 0.424 | | N | -0.633 |
| | N | -0.505 | | C$_N$ | -0.293 |
| | H$_N$ | 0.114 | | H$_{CN}$ | 0.130 |
| **BFNF** | F$_B$ | -0.243 | **B(NH$_2$)N(NH$_2$)** | H$_{NB}$ | 0.231 |
| | B | 0.820 | | H$_{NB}$ | 0.237 |
| | N | -0.383 | | N$_B$ | -0.622 |
| | F$_N$ | -0.194 | | B | 0.704 |
| **B(OH)N(OH)** | H$_B$ | 0.339 | | N | -0.562 |
| | O$_B$ | -0.521 | | N$_N$ | -0.469 |
| | B | 0.748 | | H$_{NN}$ | 0.246 |
| | N | -0.469 | | H$_{NN}$ | 0.237 |
| | O$_N$ | -0.437 | **B(CHO)N(CHO)** | H$_B$ | 0.082 |
| | H$_N$ | 0.340 | | O$_B$ | -0.165 |
| **B(CN)N(CN)** | N$_B$ | -0.029 | | C$_B$ | 0.044 |
| | C$_B$ | -0.019 | | B | 0.536 |
| | B | 0.608 | | N | -0.616 |
| | N | -0.646 | | C$_N$ | 0.212 |
| | C$_N$ | 0.141 | | O$_N$ | -0.207 |
| | N$_N$ | -0.055 | | H$_N$ | 0.114 |

Figure captions (color online):

Fig-1: Total / site projected density of states and band structure of (a) BN sheet and BN sheet functionalized by (b) H, (c) F, (d) OH, (e) CHO, (f) $CH_3$, (g) CN and (h) $NH_2$.

Fig-2: Phonon dispersion and density of states plots of (a) BN sheet and BN sheet functionalized by (b) H, (c) F, (d) OH, (e) CHO, (f) $CH_3$, (g) CN and (h) $NH_2$.

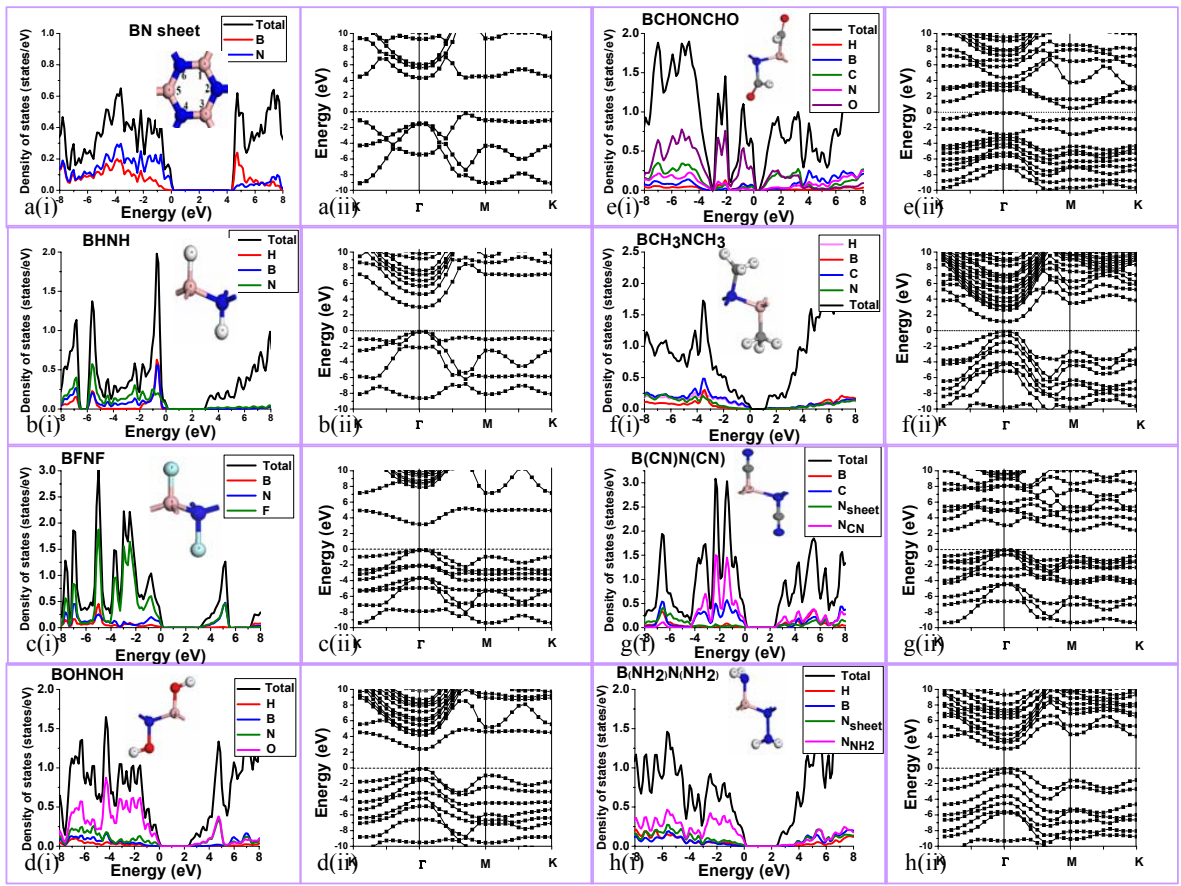

Fig-1

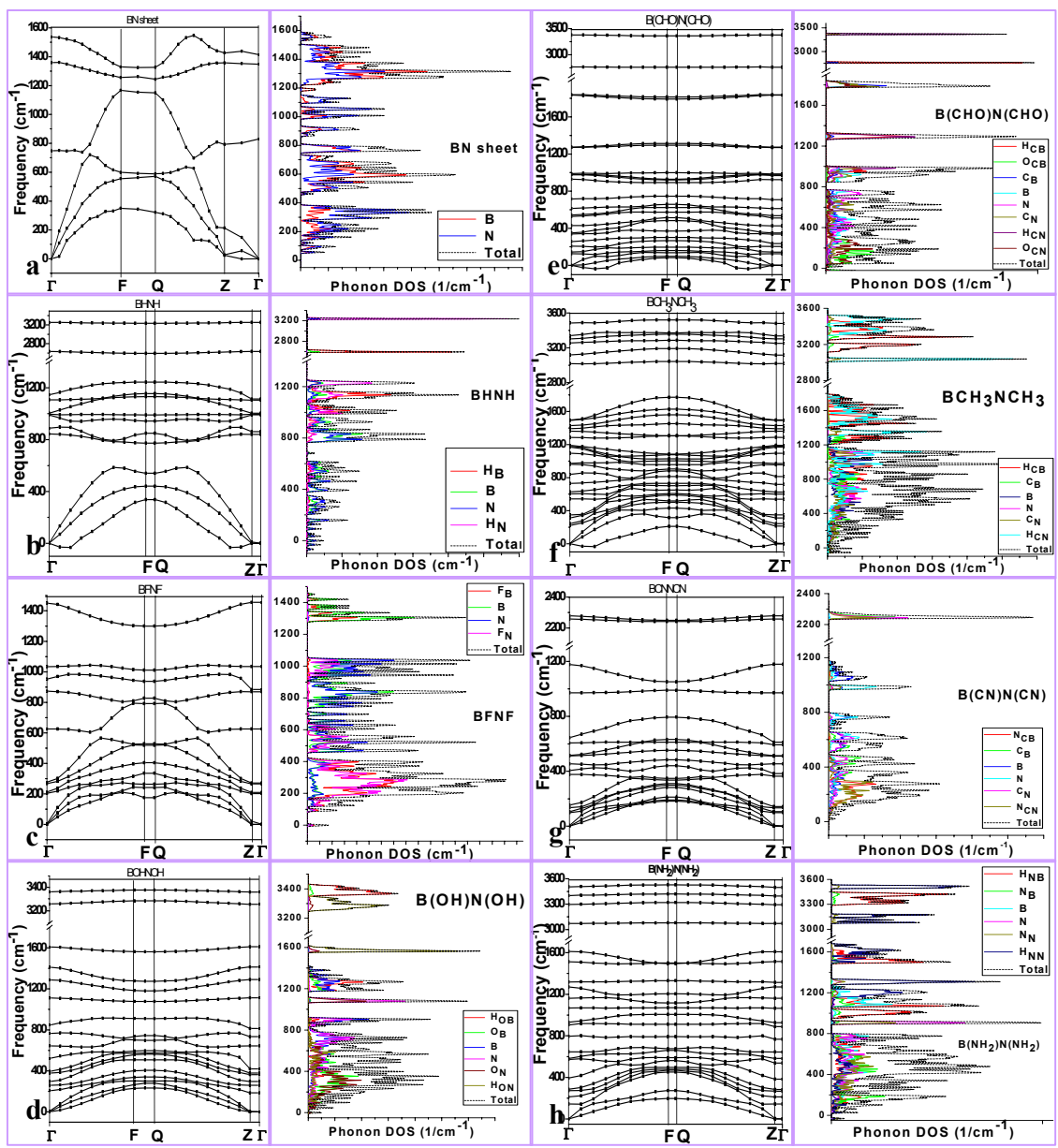

Fig-2